\documentclass[journal]{IEEEtran}
\usepackage{cite}
\usepackage{amsmath,amssymb,amsfonts}
\usepackage{graphicx}
\usepackage{textcomp}
\usepackage{stackengine}[2021-07-15]
\usepackage[dvipsnames]{xcolor}
\usepackage[caption=false,font=normalsize,labelfont=sf,textfont=sf]{subfig}
\usepackage{epsfig,rotating,setspace,latexsym,amsmath,epsf,amssymb,amsfonts,bm,theorem,cite,graphicx,epsf,authblk,epstopdf,url,color,mwe,xcolor,multirow, booktabs,threeparttable}
\usepackage{amsmath,amsfonts}
\usepackage{algorithm}
\usepackage{algpseudocode}
\usepackage{array}
\usepackage{gensymb}
\usepackage{textcomp}
\usepackage{stfloats}
\usepackage{url}
\usepackage{verbatim}
\usepackage{graphicx}

\begin{document}
\title{Area and Power Efficient FFT/IFFT Processor for FALCON Post-Quantum Cryptography}

\author{Ghada~Alsuhli,
        Hani~Saleh,~\IEEEmembership{Member,~IEEE,}
        Mahmoud~Al-Qutayri,~\IEEEmembership{Member,~IEEE,}
        Baker~Mohammad,~\IEEEmembership{Member,~IEEE,}
        and~Thanos~Stouraitis,~\IEEEmembership{Life~Fellow,~IEEE}}



\maketitle

\begin{abstract}
Quantum computing is an emerging technology on the verge of reshaping industries, while simultaneously challenging existing cryptographic algorithms. FALCON, a recent standard quantum-resistant digital signature, presents a challenging hardware implementation due to its extensive non-integer polynomial operations, necessitating FFT over the ring $\mathbb{Q}[x]/(x^n+1)$. This paper introduces an ultra-low power and compact processor tailored for FFT/IFFT operations over the ring, specifically optimized for FALCON applications on resource-constrained edge devices. The proposed processor incorporates various optimization techniques, including twiddle factor compression and conflict-free scheduling. In an ASIC implementation using a 22 nm GF process, the proposed processor demonstrates an area occupancy of 0.15 mm$^2$ and a power consumption of 12.6 mW at an operating frequency of 167 MHz. Since a hardware implementation of FFT/IFFT over the ring is currently non-existent, the execution time achieved by this processor is compared to the software implementation of FFT/IFFT of FALCON on a Raspberry Pi 4 with Cortex-A72, where the proposed processor achieves a speedup of up to 2.3$\times$. Furthermore, in comparison to dedicated state-of-the-art hardware accelerators for classic FFT, this processor occupies 42\% less area and consumes 83\% less power, on average. This suggests that the proposed hardware design offers a promising solution for implementing FALCON on resource-constrained devices. 

\end{abstract}

\begin{IEEEkeywords}
Post-quantum cryptography, FALCON, ASIC, FFT/IFFT processor, Polynomial operations.
\end{IEEEkeywords}

\IEEEpeerreviewmaketitle

\section{Introduction}
\label{sec:introduction}
Cryptography is crucial for ensuring the security of information processing and communication, especially for applications that involve sensitive information, such as online banking, medical devices, and autonomous cars. However, the development of quantum algorithms, like Shor’s
and Grover’s
, has shown that commonly used symmetric and asymmetric key cryptographic schemes, such as the Rivest-Shamir-Adleman (RSA) algorithm, can be easily broken by powerful quantum computers \cite{chamola2021information}. This means that the integrity and confidentiality of information communications could be seriously threatened once such computers become widely available. Several leading companies, including IBM, Intel, and Google, are currently working on developing superconducting quantum processors \cite{deshpande2022assessing}. Although these quantum computers are not yet powerful enough to pose a threat, they represent a significant step toward the development of more powerful quantum technology in the future.

To prepare for the post-quantum era, a new round of cryptosystem innovation has recently been initiated and become an active research topic \cite{bernstein2017post}. In addition, the National Institute of Standards and Technology (NIST) has launched a post-quantum standardization process for standardizing new Post-Quantum Cryptography (PQC) algorithms that remain secure even in worst-case scenarios when an attacker has a quantum computer. As a result of this standardization process, several   Digital Signature (DS) and Key Encapsulation Mechanism (KEM) cryptosystems that are believed to be quantum-resistant have been identified and selected for standardization, such as SPHINCS, CRYSTALS-KYBER, CRYSTALS-Dilithium, and FALCON. 

As the NIST PQC standardization process approaches its end, the deployment of PQC algorithms and the evaluation of their hardware efficiency have become major concerns in cryptographic engineering \cite{joseph2022transitioning}. Many research studies have focused on building dedicated hardware for PQC algorithms \cite{he2023hardware,imran2023high,canto2022error,lucas2022lightweight,shahbazi2023optimized}, aiming either to implement vanilla PQC algorithms \cite{he2023hardware,imran2023high,canto2022error} or their lightweight versions \cite{lucas2022lightweight, shahbazi2023optimized}. However, the literature is more focused on a few PQC algorithms, such as CRYSTALS-KYBER and CRYSTALS-Dilithium. This narrow focus is attributed to the simple approach of CRYSTALS algorithms (both based on Learning with Errors (LWE)) which has been studied widely even before the standardization process started. However, other algorithms, such as the FALCON DS scheme \cite{Fouque2019FalconFL}, have some advantages over CRYSTALS algorithms.

FALCON is a digital signature algorithm known for its quantum security and efficiency in terms of communication bandwidth and verification simplicity. It is expected to be a popular choice for IoT applications due to its smaller signature and public key size and faster verification process compared to other signature schemes like CRYSTALS–Dilithium \cite{alagic2022status}. However, FALCON is not naturally hardware-friendly, and this poses a challenge for implementation on resource-constrained devices typically found in IoT scenarios. These devices have low power budgets, limited computation capabilities, small memory, and/or low communication bandwidth. Any hardware implementation of FALCON should be designed to take into account these constraints.

      

FALCON consists of three main stages: key generation, signing, and verification. The key generation and signing heavily rely on Fast Fourier Transform (FFT) calculations. In fact, FFT accounts for 26\% and 48\% of the total clock cycles at key generation and signing processes, respectively \cite{kim2022accelerating}. Additionally, to claim meaningful security bounds for FALCON, FFT with double-precision Floating-Point (FP) arithmetic is required \cite{Fouque2019FalconFL}. However, this poses a significant limitation for resource-constrained devices that lack a Floating-Point Unit (FPU). Moreover, the FFT accelerator should be designed to support all possible FALCON standard parameters, with the main FFT-related parameter being $N \in\{512, 1024\}$, which defines the dimension of the lattice. Even with specific parameters, FALCON uses FFTs of different sizes, which are determined by the parameters of FALCON and are powers of two that divide $N$. 
\begin{figure}[ht]
    \centering    
    \includegraphics[width= \linewidth]{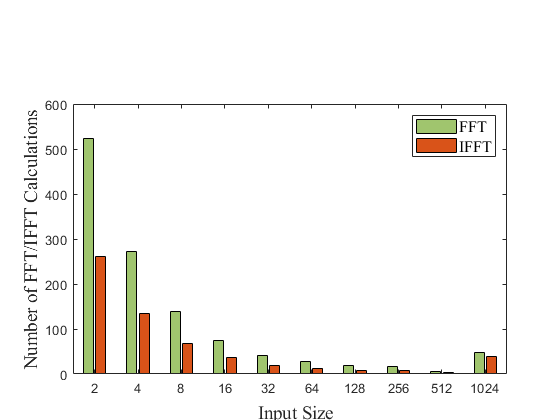}
    \caption{Required FFT/IFFT calculations and the corresponding sizes for FALCON, for security parameter $N=1024$}
    \label{fft_op}
      
\end{figure}

Figure \ref{fft_op} illustrates the number of required FFT/IFFT per FFT size for FALCON parameter $N=1024$. It can be observed from this figure that (i) hundreds of FFT/IFFT calculations are required, and (ii) FALCON needs more FFT/IFFT calculations of small sizes. Furthermore, it should be noted that, unlike the typical FFT/IFFT, FALCON uses FFT/IFFT over the ring $\mathbb{Q}[x]/(\phi)$, where $\phi$ is the irreducible polynomial used in the ring and $\mathbb{Q}$ is the set of all rational numbers. Consequently, custom FFT algorithms are used to implement FFT and Inverse FFT (IFFT) \cite{Fouque2019FalconFL}. Therefore, a customized, high-precision, and on-the-fly configurable FFT accelerator that considers different requirements of FALCON while taking into account the constraints of resource-constrained devices is required. 

The contributions of this paper are summarized as follows:

\begin{itemize}
\item We have examined the specific requirements and unique characteristics of FFT/IFFT over a ring for FALCON PQC. To implement an efficient hardware solution, we utilize the special properties of FFT/IFFT over the ring and adopt an approach inspired by the Negative Wrapped Convolution (NWC) method \cite{lyubashevsky2008swifft}. This implementation is, to our knowledge, the first of its kind in the literature.

\item We propose a Hardware architecture of the processor that relies on a processing elements (PE) array, distributed single-port SRAMs, and distributed ROMs. Each component of the architecture is optimized for power and area efficiency. The proposed architecture offers three levels of reconfigurability. It considers the security parameters, accommodates various input sizes based on the polynomial degree and supports both FFT and IFFT operations.

\item We present a conflict-free scheduling algorithm that enables multiple PEs to access the coefficient memory simultaneously without conflicts. This approach maximizes the utilization of the PEs and ensures stalling-free memory access.

\item We have compressed the size of the twiddle factor table used in the FALCON reference implementation from 16 KB to just 4 KB for more area and power reduction.

\item Using the Global Foundries (GF) 22 nm process, our Application-Specific Integrated Circuit (ASIC) implementation occupies an area of 0.15 mm$^2$ and consumes 12.6 mW of power at an operating frequency of 167 MHz. Given the absence of hardware implementations for similar processors, our processor exhibits a noteworthy 2.3$\times$ speedup compared to the software implementation of FFT/IFFT in FALCON on a Raspberry Pi 4 with Cortex-A72. Additionally, when assessed against state-of-the-art classic FFT processors in terms of normalized area and power, our design demonstrates a 42\% improvement in area efficiency and an 83\% enhancement in power efficiency on average. These findings firmly establish our proposed hardware design as an encouraging solution for implementing FALCON on resource-constrained devices.

\end{itemize}

\section{Literature Review}
In this section, we discuss the available work related to optimized implementations of FALCON and the state-of-the-art FFT/IFFT accelerators.
\subsection{FALCON Implementation Optimization}
Several implementations of the FALCON PQC algorithm that aim to optimize its performance for different platforms and applications have been proposed. In addition to the official reference implementation of FALCON available for multiple platforms, including ARM Cortex-M4 and x86-64 processors \cite{alagic2022status}, there are several optimized software implementations \cite{kim2022accelerating, nguyen2022fast, oder2019towards}.

No full hardware implementation of FALCON exists to date. Available hardware solutions consider implementing the verification part only, which requires only integer, rather than rational, polynomial operations. 
Because the key and signature generations of FALCON are delicate to implement, the authors in \cite{beckwith2022high} chose to implement the hardware of the verification part only. The main focus of this work was to optimize the polynomial multiplication in the integer domain using the Number Theoretic Transform (NTT)\footnote{NTT can be seen as an FFT in a finite field, in which polynomial arithmetic takes place in the integer domain.} and optimize its hardware architecture. Similarly, Field-Programmable Gate Arrays (FPGA) and ASIC implementations and Optimizations for the verification process using High-Level Synthesis (HLS) have been proposed \cite{soni2021hardware}. Since FALCON key generation and signing parts require FP operations and recursive functions, the authors could not use HLS to generate the hardware of these parts. In addition, a hardware-software co-design approach was employed to implement FALCON verification. This resulted in an architecture that incorporates a RISC-V processor integrated with various hardware accelerators, mainly performing the efficient execution of the hash function and polynomial multiplications \cite{karl2022post}. 

The lack of full hardware implementation of FALCON is attributed to the facts that: (i) FALCON is non-trivial to understand and needs a rigid understanding of the underlying mathematical and security concepts \cite{oder2019towards}, (ii) FALCON is a relatively new PQC algorithm, (iii) FALCON relies heavily on FFT over the ring calculations in the complex domain which require double-precision FP arithmetic 
to be used. In this work, we target designing a hardware accelerator for FFT over the ring which is the main expensive operation of FALCON key generation and signing processes. The integration of the proposed accelerator with a FALCON System on a Chip (SoC) or within the pipeline of a FALCON crypto-processor is anticipated to facilitate the development of a complete hardware or hardware-software co-design implementation for FALCON.   

\subsection{FFT Accelerators}
Hardware acceleration for FFT has been the subject of extensive research over the past few decades \cite{garrido2022survey}. Numerous FFT hardware accelerators have been proposed in the literature \cite{Bertaccini2021buffer, cilasun2020crafft, di2021multi}. With the growing proliferation of FFT applications, there has been a trend towards customized hardware designs tailored for specific use cases \cite{Lee20202,liu2018high}, while others aim for a more generic solution to accommodate the requirements of different applications \cite{Xu2022scalable,chen2018variable,  wang20170}. For instance, several recent accelerator designs involve reconfigurability to enable FFT computations on variable size input \cite{xia2017memory, Xu2022scalable,chen2018variable}.
In the case of FALCON, it becomes imperative to include such reconfigurability. As depicted in Figure \ref{fft_op}, FALCON scheme necessitates extensive FFT operations on polynomials of varying degrees, indicating the need for FFTs with different sizes. 

FFT accelerators commonly employ fixed-point, i.e. integer, arithmetic for calculations \cite{ li2022fixed, saleh2013high, sun2022approximating}. However, this approach introduces approximation and low precision to the FFT computations. In contrast, some researchers have opted for single-precision FP (32-bit) representations to enhance FFT accuracy \cite{chen2018variable,wang2016pipelined,wang2019area,beulet2014low }. Although the use of FP arithmetic reduces FFT errors, it comes at the cost of reduced speed, increased power consumption, and greater chip silicon requirements \cite{chen2018variable}. When it comes to FALCON, the double-precision FP format (64-bit) is essential for proper functionality \cite{Fouque2019FalconFL}. This requirement adds further complexity to the design of an FFT specifically tailored for FALCON.

Utilizing FFT for polynomial multiplication has become prevalent in security applications, including homomorphic encryption  \cite{duong2022area,  feng2018accelerating} and post-quantum cryptography \cite{mu2022scalable, zhang2022ultra}. These works target the multiplication of polynomials over the ring of integers, in which the FFT becomes an NTT. Unlike other PQC algorithms, FALCON requires polynomial operations to happen frequently in both complex and integer fields, i.e. FFT-based and NTT-based polynomial operations. In this work, we focus on implementing FFT-based polynomial operations in the complex domain. To the best of our knowledge, such an implementation has not yet been presented in the existing literature. 

\section {Preliminaries}

This section focuses on key concepts and algorithms relevant to the classic FFT/IFFT as well as the FFT/IFFT over a ring, which is required for FALCON. Additionally, it discusses the utilization of FFT/IFFT over a ring for polynomial operations in FALCON and explores how to implement it efficiently. 

\subsection{Basics of Classic FFT/IFFT}
FFT is an efficient algorithm for computing the Discrete Fourier Transform (DFT) of a sequence of $n$ numbers $\boldsymbol{a} \in \mathbb{R}^n$ ($\mathbb{R}$ is the set of real numbers). The classic DFT is defined as 
$\hat{a}_k = FFT (\boldsymbol{a}) = DFT (\boldsymbol{a})= \sum_{j=0}^{n-1} a_j \omega _n^{kj}$, where $k = 0, 1, ..., n-1$ and $\omega_n= e^{2\pi i/n}$ is a primitive $n$th root of $1$ in the complex domain $\mathbb{C}$, also known as the twiddle factors. The classic inverse DFT is obtained by replacing $\omega _n$ with $\omega _n^{-1}$ and normalizing the result by $n$, i.e. ${a}_j = IFFT (\boldsymbol{\hat{a}})= IDFT (\boldsymbol{\hat{a}})= \frac{1}{n}\sum_{k=0}^{n-1} \hat{a}_k \omega _n^{-kj}$,where $j = 0, 1, ..., n-1$.

\begin{figure*}[ht]
    \centering
    \includegraphics[width=0.7\textwidth]{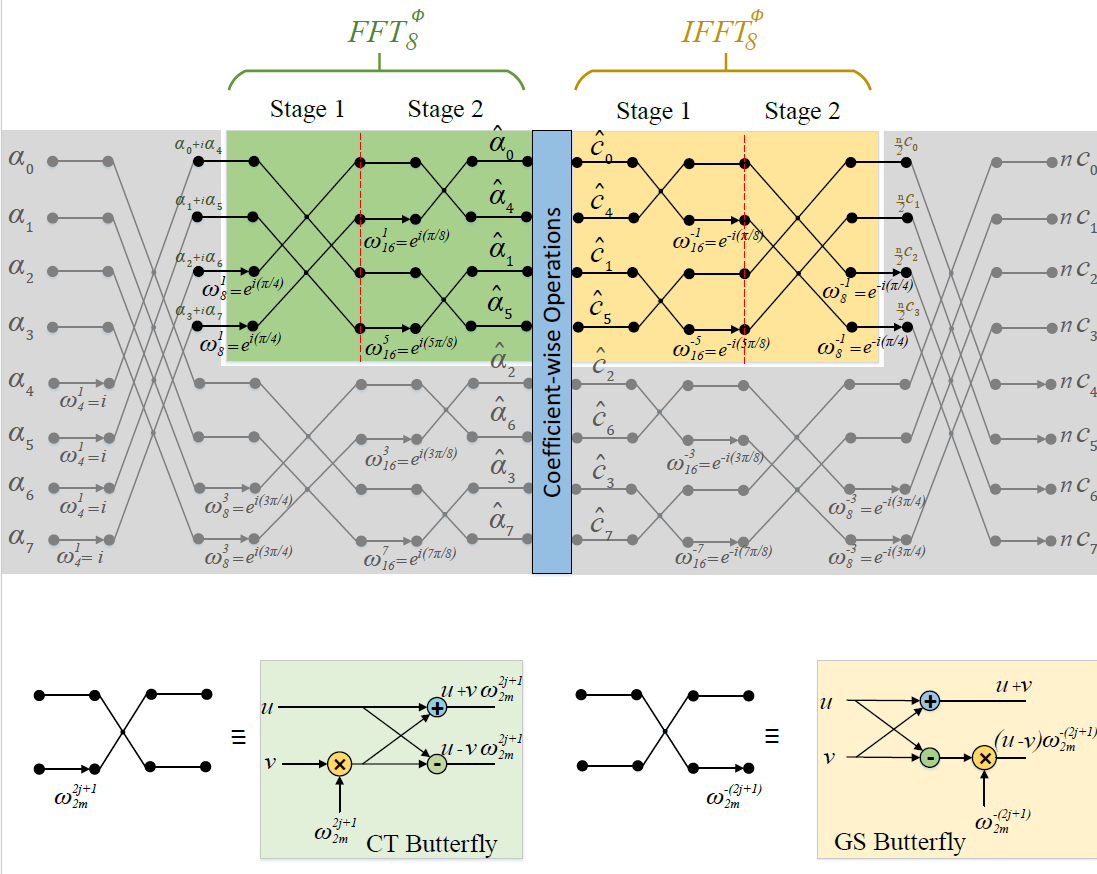}
    \caption{Dataflow of $FFT^\phi_8$ and $IFFT^\phi_8$ used for FALCON. The coefficient-wise operations can be any operation to be executed efficiently in FFT representation such as polynomial multiplication and division. The lower part of this figure shows the CT and GS butterflies used for $FFT^\phi$ and $IFFT^\phi$, respectively. }
    \label{FFT_flow}
      
\end{figure*}

FFT can be calculated using several algorithms. The most commonly used algorithm is the Cooley-Tukey (CT) algorithm which relies on the divide-and-conquer strategy to recursively divide every DFT into two smaller DFTs until we reach 2-point DFTs, which are called butterflies. This is usually called Decimation-in-Time (DIT) FFT. Another common algorithm is the Decimation-in-Frequency (DIF) or Gentleman-Sande (GS) algorithm which uses different butterflies. The two butterflies used in these algorithms are illustrated in the lower part of Figure \ref{FFT_flow}.

\subsection{FFT-based Polynomial Operations}

In FALCON, polynomials are defined over the ring $\boldsymbol{R}[x] = \mathbb{Q}[x]/(\phi)$, where $\phi= x^n + 1$ is a cyclotomic polynomial and $n$ is a power of $2$. FALCON also presents polynomials over the quotient ring $\mathbb{Z}/q\mathbb{Z}$ with prime $q$. Polynomial over the ring $\boldsymbol{R}[x]$ are defined by $\boldsymbol{a}=\sum_{i=0}^{n-1} a_i x^i$, where $a_i \in \mathbb{Q}$ are the $i$th coefficients of the polynomial $\boldsymbol{a}$. Performing additions, subtractions, multiplications, and divisions of polynomials modulo $\phi$ can be efficiently computed in FFT representations by applying the corresponding operations to each coefficient of the polynomials, i.e. coefficient-wise operations. Next, we investigate using FFT for polynomial multiplication in more detail.

Let $\boldsymbol{a}$ and $\boldsymbol{b}$ be two polynomials of the ring $\boldsymbol{R}[x] $. Using the convolution in the frequency domain, the product $\boldsymbol{c} \in \boldsymbol{R}[x] $ of these two polynomials can be calculated using the classic FFT/IFFT as 
\begin{equation} \label{FFT_mul}
\boldsymbol{c}= \boldsymbol{a} \times \boldsymbol{b} = [IFFT_{2n}(FFT_{2n}(\boldsymbol{a})\otimes FFT_{2n}(\boldsymbol{b}))] \text{ mod } \phi.
\end{equation}
where $\otimes$ denotes point-wise multiplication and the notation $2n$ in $FFT_{2n}$($ IFFT_{2n}$) indicates the FFT(IFFT) size. Namely, the polynomial multiplication over $\boldsymbol{R}[x] $ can be calculated by applying $2n$-point FFT on the polynomials after padding them, performing coefficient-wise multiplication, applying $2n$-point IFFT on the product, and finally reducing the resultant polynomial to the degree of $n-1$ with the modular polynomial $\phi$.  

For more efficient exploitation of FFT for polynomial multiplication over $\boldsymbol{R}[x]$, we adopt an approach inspired by NWC method \cite{lyubashevsky2008swifft} that is recently applied to NNT \cite{zhang2019nttu,mu2022scalable}. This approach allows the application of $n-$point instead of $2n-$point FFT/FFT and alleviates the need to apply the modular reduction. With NWC, the polynomial multiplication in $\boldsymbol{R}[x] $ becomes    
\begin{equation} \label{FFT_mul1}
\boldsymbol{\dot{c} }= IFFT_{n}(FFT_{n}(\boldsymbol{\dot{a}})\otimes FFT_{n}(\boldsymbol{\dot{b}})),
\end{equation}
where $\boldsymbol{\dot{a} }=\sum_{i=0}^{n-1} \psi_{2n}^{i} a_i x^i$ and $\boldsymbol{\dot{b} }=\sum_{i=0}^{n-1} \psi_{2n}^{i} b_i x^i$ are the two polynomials $\boldsymbol{{a} }$ and $\boldsymbol{{b} }$ after applying pre-processing, i.e. scaling, to them. The used scaling factor $\psi_{2n}$ is the $2n$th root of unity with $\psi_{2n}^2=\omega _n$. Similarly, the actual product is calculated by applying post-processing such that $\boldsymbol{{c} }=\sum_{i=0}^{n-1} \psi_{2n}^{-i} \dot{c}_i x^i$.

The vanilla NWC requires $3\times n$ extra complex multiplications for FFT/IFFT-based polynomial multiplication, which is expensive for large $n$. Fortunately, it has been shown that this scaling can be merged with the twiddle factor when NTT is implemented using the Cooley-Tukey algorithm  \cite{longa2016speeding}. Similarly, merging the scaling for INNT is possible and requires switching to the Gentleman-Sande algorithm \cite{zhang2020highly}. Following the same approach, our FFT and IFFT are implemented using two separate algorithms and two different butterflies, to let the twiddle factors absorb the pre-processing and post-processing overheads. 

Using the notations $FFT^\phi$ and $IFFT^\phi$ to distinguish FFT/IFFT over $\mathbb{Q}[x]/(\phi)$ from the classic FFT/IFFT, 
the FFT of a polynomial $\boldsymbol{{a} }$ after applying the NWC becomes
\begin{flalign} \label{FFT_NWC}
\hat{a}_k = FFT^\phi_n (\boldsymbol{a})&=\sum_{j=0}^{n-1} a_j \psi_{2n}^{j} \omega _n^{kj}&\\ \nonumber
&=\sum_{j=0}^{n-1} a_j \omega _{2n}^{j(2k+1)}, \nonumber
\end{flalign}
while the IFFT is
\begin{flalign} \label{IFFT_NWC}
{a}_j = IFFT^\phi_n (\boldsymbol{\hat{a}})&= \frac{1}{n}\sum_{k=0}^{n-1} \hat{a}_k \psi_{2n}^{-k} \omega _n^{-kj}&\\ \nonumber
&= \frac{1}{n}\sum_{k=0}^{n-1} \hat{a}_k \omega _{2n}^{-k(2j+1)}.\nonumber
\end{flalign}
It is worth noting that $\omega (k)=\omega_{2n}^{2k+1}=e^{i(\pi/n)(2k+1)}, \text{ } k=0,..,n-1$, are the complex roots of $\phi$ over $\mathbb{C}$ ($n$ distinct roots). Thus, calculating the FFT of a polynomial $\boldsymbol{a } \in \mathbb{Q}[x]/(\phi)$ using NWC is equivalent to evaluating it for the complex roots of $\phi$ over $\mathbb{C}$, i.e. calculating $\boldsymbol{{a}} (\omega (k))$.

By applying the divide-and-conquer approach followed by the Cooley-Tukey algorithm, which divides the coefficients based on their index parity, for $FFT^\phi$ in equation (\ref{FFT_NWC}), we get 
\begin{flalign} \label{FFT_NWC_CT}
\hat{a}_k &=\sum_{j=0}^{n/2-1} a_{2j}\  \omega _{n}^{j(2k+1)}&+&\omega _{2n}^{(2k+1)}\sum_{j=0}^{n/2-1} a_{2j+1}\  \omega _{n}^{j(2k+1)},\\ \  \nonumber
&=\hat{a}_k^1&+&\omega _{2n}^{(2k+1)}\ \hat{a}_k^2,\\ \nonumber
\hat{a}_{k+n/2}&=\hat{a}_k^1&-&\omega _{2n}^{(2k+1)}\ \hat{a}_k^2, \nonumber
\end{flalign}
where $k= 0,1,...,n$, $\hat{a}_k^1= \sum_{j=0}^{n/2-1} a_{2j}\ \omega _{n}^{j(2k+1)}$ and $\hat{a}_k^2=\sum_{j=0}^{n/2-1} a_{2j+1}\  \omega _{n}^{j(2k+1)}$ are the $FFT^\phi_{n/2}$ of $a_{2j}$ and $a_{2j+1}$, respectively. This decimation is applied recursively until we reach a $FFT^\phi_2$. Similarly, the Gentleman-Sande algorithm can be applied to (\ref{IFFT_NWC}) for the IFFT case.  Figure \ref{FFT_flow} shows the data flows of $FFT^\phi_8$ and $IFFT^\phi_8$ as an example.

$\omega (k)$ has several properties that further simplify the calculation of $FFT^\phi$ and $IFFT^\phi$. First, $\omega(n-1-k) = \text{{conjugate}} (\omega(k)) = 1/\omega (k)$, where $\text{{conjugate}} (x)$ is the complex conjugate of $x$. Given that the polynomial $\boldsymbol{{a}} \text { is in }\mathbb{Q}[x]/(\phi)$, it follows that $\boldsymbol{{a}}(\omega(n-1-k)) = \text{{conjugate}}(\boldsymbol{{a}}(\omega (k))$. As a result, it is sufficient to evaluate $\boldsymbol{{a}}(\omega (k))$ for $k = 0,...,n/2-1$, which reduces the storage space and calculation requirements for $FFT^\phi/IFFT^\phi$ by a factor of $2$ \cite{Fouque2019FalconFL}. The remaining $\boldsymbol{{a}}(\omega (k))$ for $k = n/2,...,n-1$ can be easily calculated, if necessary. 

The second useful property of $\omega (k)$ is that when the final recursion stage is reached, the twiddle factor to be evaluated is $\omega _{4}^{1} =i$. Hence, we can think about the first stage of FFT flow as transforming the FFT input $\in \mathbb{Q}$ into the complex domain such that the first stage can be eliminated and the input of the second stage is $a^{(0)}_{k}=a_{k}+i\ a_{k+n/2}$ for $k=0,...,n/2-1$. This step is easily done in hardware by separating the storage of the imaginary and real parts and storing the second $n/2$ of the input as an imaginary part. The discussion above is also applicable to the case of IFFT. We should mention here that to consider the impact of the ignored redundant half of the FFT in the final stage of IFFT, the real and imaginary values in the output should be duplicated. This is compensated by normalizing the IFFT output over $n/2$ instead of $n$.

Consequently, we can think about $FFT^\phi_n/IFFT^\phi_n$ as an FFT/IFFT of $n/2$ complex numbers with special twiddle factors. This reduces the size of FFT by a factor of 2, as illustrated in Figure \ref{FFT_flow}, where the grey part of the dataflow shows the reduced unnecessary calculation because of exploiting the two properties of $\omega(k)$ mentioned above. The $FFT^\phi_n/IFFT^\phi_n$ algorithm consists of $\log_2(n)-1$ stages with each stage involving $n/4$ butterfly operations.

Figure \ref{FFT_flow} reveals that the order of stage 2 in FFT (and stage 1 in IFFT) differs from the natural order derived in equation (\ref{FFT_NWC_CT}). This variation in the ordering is due to FALCON's original software implementation \cite{falconsign}, where this specific order was chosen to simplify the software computations. While this ordering produces different results, it functions correctly as long as the same order is maintained for both FFT and IFFT \cite{Fouque2019FalconFL}. Although our hardware implementation does not require this order, we adopt the same ordering for easier verification of our hardware design by comparing it to the FALCON software implementation.


One problem with using CT FFT and GS IFFT is that the CT usually requires the FFT input to be in bit-reversed \footnote{The bit-reversed order refers to the rearrangement of the input sequence of data points, swapping their positions in a binary reflection pattern.} order and produce the FFT output in the natural order. On the other hand, the opposite happens with GS IFFT. Thus, the bit-reversing step is needed before FFT and after IFFT when calculating the polynomial operations. To overcome this, we follow an approach similar to the one observed for NTT in \cite{liu2017high}, in which the CT and GS algorithms are manipulated and the butterflies are used independently, as illustrated in Figure \ref{FFT_flow}. This alleviates the need for any bit-reversal calculation at the beginning or end of the polynomial operation calculation. Further manipulation of these algorithms is needed in our processor to obtain conflict-free scheduling, as we will see in Section \ref{Scheduling}. 

\section{DESIGN METHODOLOGY}
In this section, we describe the design methodology of our processor hardware implementation. Firstly, we explain the runtime reconfigurability of our processor to accommodate the requirements of FALCON. Secondly, we provide an in-depth discussion of the architecture adopted for our processor. Furthermore, we discuss the various optimization techniques employed in our design, including twiddle factor preprocessing and conflict-free scheduling.

\subsection{Reconfigurability of $FFT^\phi$/$IFFT^\phi$ Processor}
The proposed design of the $FFT^\phi$/$IFFT^\phi$ processor for FALCON incorporates three levels of reconfigurability.

Firstly, the design takes into account all security parameters that impact the processor design. The primary parameter of interest is the lattice dimension $N$, which determines the quantum hardness of FALCON. Our processor is designed to support both security levels of FALCON, namely NIST Level I and V, which require $N$ to be either 512 or 1024, respectively. As a result, the maximum size of $FFT^\phi$/$IFFT^\phi$ supported by our processor is 1024.

The second level of reconfigurability arises from the key generation process in FALCON, which involves calculations such as NTRU equations and fast Fourier sampling. These calculations begin with a polynomial of degree 512 or 1024, depending on the security level, and the polynomial's degree is halved iteratively until a single rational coefficient is obtained \cite{kim2022accelerating}. Consequently, our processor is optimized to dynamically adapt to various FFT sizes $FFT^\phi_{n}$/$IFFT^\phi_{n}$ with $\log(n) \in \{1,2,...,\log(N)\}, N\in\{512, 1024\}$.

Thirdly, the proposed processor is designed to handle both $FFT^\phi$ and $IFFT^\phi$, which utilize the CT and GS algorithms. Therefore, our processor can be configured as an $FFT^\phi$ processor or an $IFFT^\phi$ processor, requiring appropriate adjustments to the control flow, memory access, and datapath.

\subsection{Proposed Overall Architecture}
\begin{figure*}[ht]
    \centering
    \includegraphics[width=\textwidth]{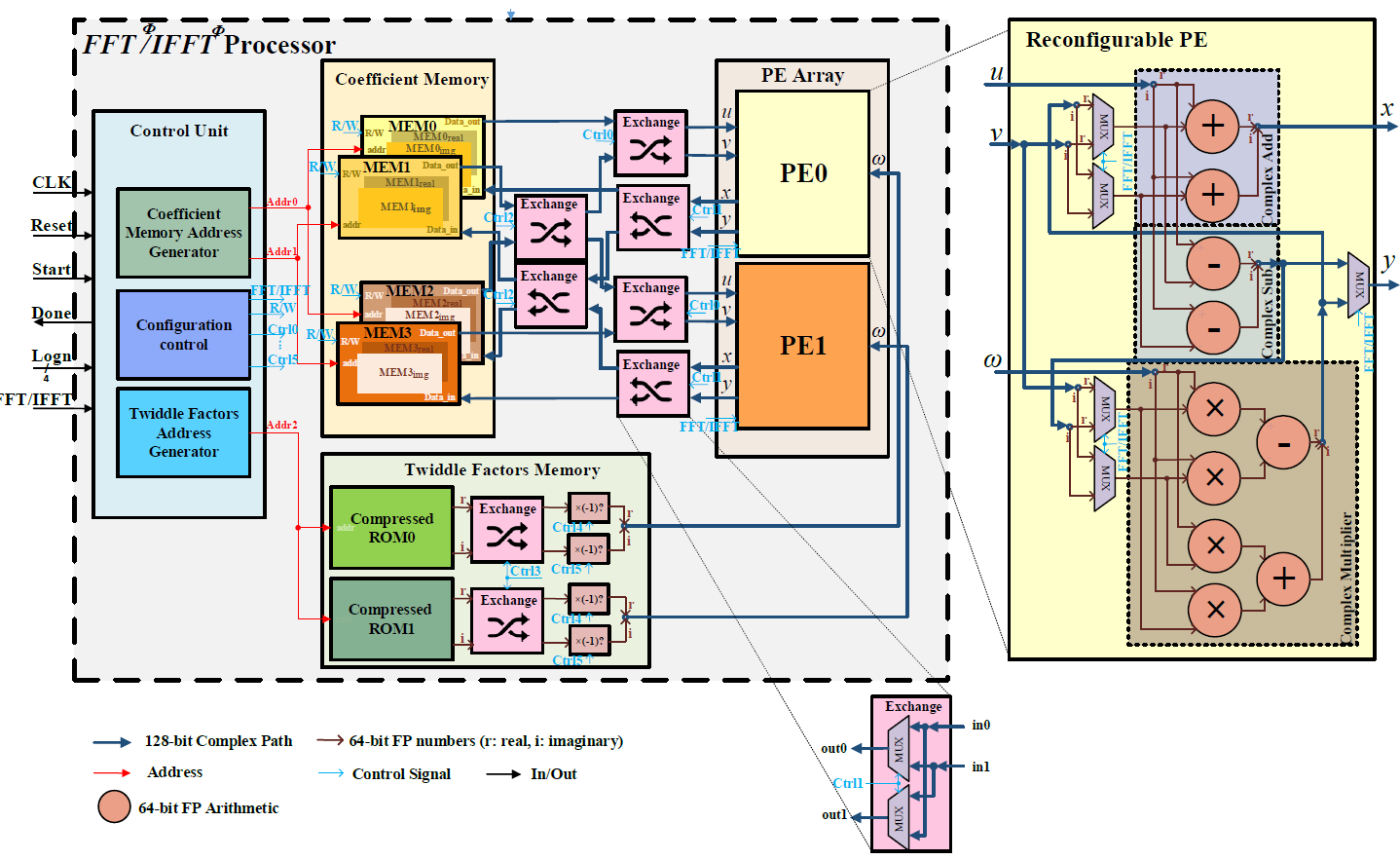}
    \caption{The architecture of the proposed $FFT^\phi$/$IFFT^\phi$ processor that involves an array of PEs, distributed coefficient memory banks, distributed twiddle factor ROMs, and a control unit. The internal design of the reconfigurable PE is illustrated on the left.}
    \label{archi}
      
\end{figure*}

We adopt an architecture that involves an array of processing elements, distributed coefficient memory banks, distributed twiddle factor read-only memory (ROM), and a control unit. An overview of the proposed architecture is illustrated in Figure \ref{archi}.

\subsubsection{Reconfigurable PE Array}

Employing a PE array architecture in an FFT accelerator brings benefits such as parallel processing, scalability for different FFT sizes, and the ability to optimize performance. These advantages make the PE array an attractive choice for our hardware implementation of $FFT^\phi$/$IFFT^\phi$ processor. When the number of PEs in the array is $n_{PE}$, the architecture is expected to perform $n_{PE}$ PE operations on every step.

Each PE in our design is adaptable depending on whether the target task of the processor is to perform $FFT^\phi$ or $IFFT^\phi$. This is determined by the $FFT/IFFT$ signal. When the signal indicates $FFT$, the PE executes radix-2 CT butterfly operations. On the other hand, when the signal indicates $IFFT$, the PE performs GS butterfly operations. Thus, the PE has three inputs $(u, v, \omega)$ and two outputs $(x, y)$ and performs radix-2 CT or GS butterfly operations.

Figure \ref{archi} shows the implementation of resource-efficient reconfigurable PE. Since CT and GS butterflies involve similar operations such as complex addition, subtraction, and multiplication, but in different orders, we utilize multiple multiplexers (MUXes) to facilitate the required reordering and reduce the overhead of duplicating hardware-unfriendly operations. The reconfigurable butterfly within the PE consists of six 64-bit FP adders/subtractors and four 64-bit FP multipliers, following the IEEE-754 double-precision FP standard. It is important to note that we have chosen not to include any form of pipelining within the PE in this work. This decision is made to avoid the overhead and complexities associated with pipeline implementation for such a wide-word (64-bit) FP arithmetic. This decision involves a trade-off between processing speed and hardware efficiency.

The choice of the number of PEs in the processor significantly influences its performance. Increasing the number of PEs can lead to faster execution for larger FFT sizes, but it comes at the expense of increased area and power consumption that are not fully utilized at all times.
Considering the characteristics of FFT operations required for FALCON and based on the analysis shown in Figure \ref{fft_op}, we observed that a substantial majority (68\%) of the required FFT operations for FALCON have relatively small sizes ($n\leq 8$). These smaller FFT sizes can be efficiently handled by a smaller number of PEs ($n_{PE}$). This allows to efficiently handle the majority of FFT operations for FALCON while minimizing the area and power consumption overhead associated with additional PEs.

\subsubsection{Distributed Coefficient Memory}

The proposed architecture utilizes multi-bank single-port SRAMs to store the initial polynomial coefficients, intermediate calculations, and the final results. Although with single-port SRAMs conflicts can arise when two PEs attempt to read from or write to the same memory bank simultaneously, it is more area and power-efficient than the dual-port one. To avoid memory read and write conflicts, it is essential to ensure that the two coefficients processed by each PE in every round are read from different memory banks and that the two output values are written to different memory banks. The minimum number of required memory banks denoted as $M$, is calculated as $M = 2\times n_{PE}$.
The size of each memory bank is given by $S_{max}/(2\times M)$, where $S_{max}=1024$ is the maximum $FFT^\phi/IFFT^\phi$ size. However, conflict is still possible unless proper conflict-free scheduling of the multi-PE is adopted. Details of this conflict-free scheduling approach will be discussed in Section \ref{Scheduling} of the paper. 

\subsubsection{Distributed Twiddle Factors Memory}
In the design of our processor, we encountered two options for implementing the twiddle factors: precomputing and storing them in memory (typically ROM), or calculating them on the fly during runtime. The first approach, involving precomputed twiddle factors, is more advantageous for the following reasons:

Firstly, the twiddle factors required for lower $FFT^\phi$ sizes are subsets of those needed for larger sizes. Therefore, we only need to store the twiddle factors required for the maximum $FFT^\phi$ size ($1024$). Furthermore, when computing the $IFFT^\phi$, we can utilize the conjugates of the twiddle factors used in the $FFT^\phi$, eliminating the need to store separate sets of twiddle factors.

Secondly, computing the twiddle factors on the fly during runtime would impose significant double-precision FP computational overhead, introducing substantial processing delays and negatively impacting the overall performance of the FALCON $FFT^\phi$/$IFFT^\phi$ processor.

Lastly, we have introduced an optimization technique that allows the reduction of the size of twiddle factor memory compared to the original implementation of FALCON. This optimization enables the minimization of the memory footprint and further enhances the efficiency of the FFT processor design. In addition, each PE uses a different set of twiddle factors which allows the distribution of the memory and the use of separate ROM for each PE. Hence, we need a total of $n_{PE}$ ROMs with a typical size of $S_{max}/(2\times n_{PE})+1$ for each ROM. These optimizations are discussed in Section \ref{twiddle}. 

\subsubsection{$FFT^\phi$/$IFFT^\phi$ Control Unit }

The controller plays a crucial role as it oversees the efficient execution of butterfly operations within the PE array. It is implemented using a Finite State Machine (FSM) and is responsible for coordinating the scheduling of operations, ensuring the optimal utilization of the PEs.
One of the primary tasks of the controller is to generate the appropriate addresses for accessing the coefficient memory banks and twiddle factor ROMs. By dynamically generating these addresses, the controller facilitates the seamless retrieval of the required data for each butterfly operation.
Furthermore, the controller is designed to be flexible and adaptable, accommodating different configurations of the processor. It can handle various operations, such as performing either $FFT^\phi$ or $IFFT^\phi$, depending on the desired functionality. Additionally, it can adapt to different input sizes, allowing the processor to handle FFT operations of varying dimensions.

\subsection{Twiddle Factors Preprocessing}\label{twiddle}

In the reference software implementation of FALCON, the twiddle factors were stored in a table using a bit-reversed order scheme, requiring a 16 KB ROM to store the twiddle factors. However, in our approach, we utilize a different indexing scheme that only requires half of the twiddle factors to be stored. This reduces the ROM storage requirement to 8 KB. Additionally, to facilitate the indexing in our conflict-free scheduling algorithm explained in the next section, we permute the order of the remaining twiddle factors to match the permutation caused by applying the conflict-free scheduling algorithm. Indeed, the CP stages of the FFT and IFFT necessitate a specific ordering of the twiddle factors.

Algorithm \ref{alg_twiddle} presents the permutation scheme used to achieve the required ordering of twiddle factors in case $n_{PE}=2$. The algorithm utilizes the function $\boldsymbol{PermutateBlocks}(x, st, sz)$, where $x$ is the vector to be processed, $st$ is the starting index of the block to be permuted, and $sz$ is the size of the block. This function swaps the specified block with the subsequent block of the same size. Specifically, for our twiddle factors, it swaps $\omega(j: j + 2^{i-3} - 1)$ with $\omega(j + 2^{i-3} : j + 2^{i-2} - 1)$. It is important to note that the preprocessing of the twiddle factors occurs offline before storing them in the ROMs, resulting in no additional overhead.

\begin{algorithm}
\caption{Twiddle Factors Permutation Algorithm} \label{alg_twiddle}
\begin{algorithmic}[1]
\Require $\omega$, $n$
\Ensure $\hat{\omega}$
\For{$i = \log_2(n) - 1$ down to $3$}
    \For{$j = 2^{i-2} + 2^i + 1$ to $n$}
        \State $\omega \gets \boldsymbol{PermutateBlocks}(\omega, j, 2^{i-3})$
        \State $j \gets j+2^{i-1}$
    \EndFor
\EndFor
\State $\hat{\omega} \gets \omega$
\end{algorithmic}
\end{algorithm}

\begin{figure*}[th!]
    \centering
    \includegraphics[width=1\textwidth]{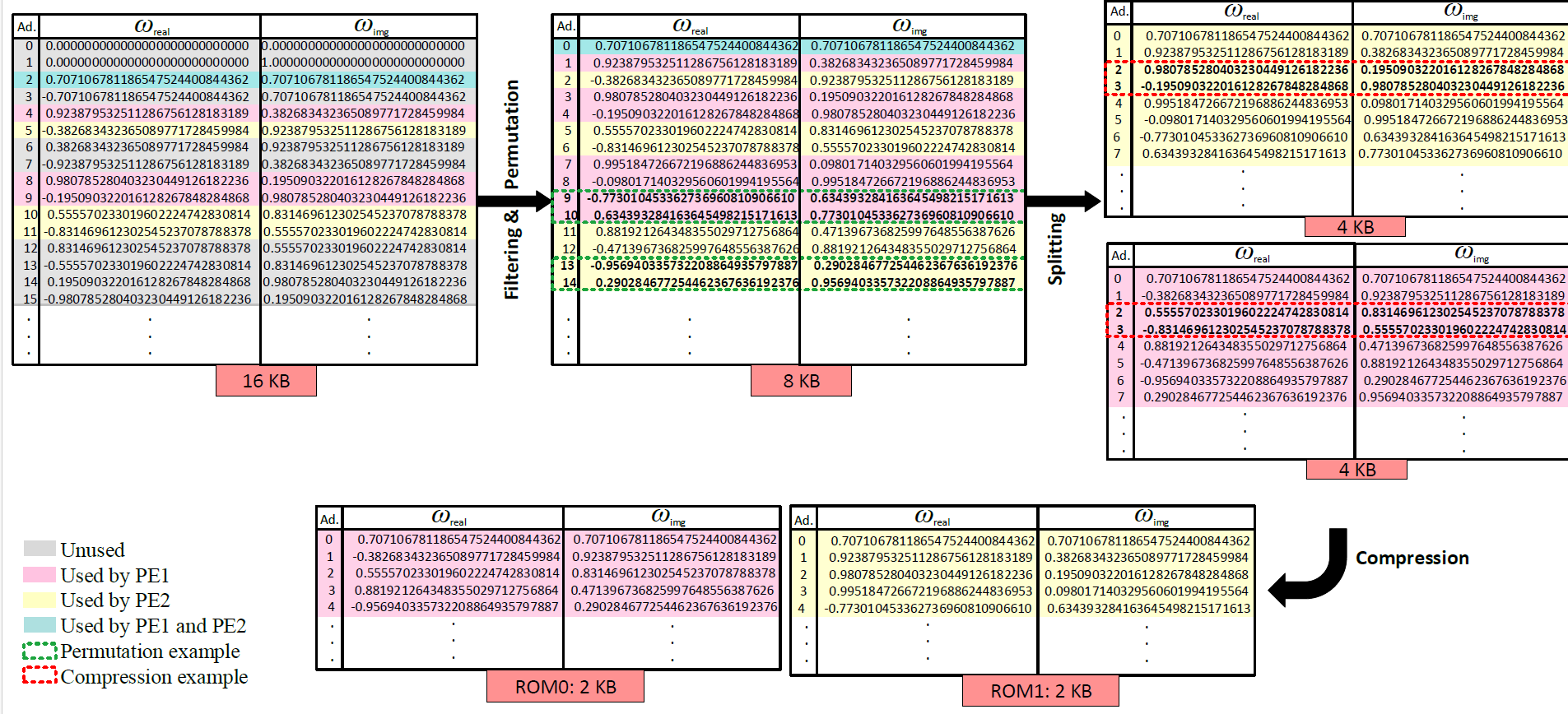}
    \caption{Twiddle factor preprocessing through filtering and permutation, along with ROM splitting and compression for the case when $n_{PE}=2$. Illustrations demonstrate instances where permutation and compression techniques are applied.}
    \label{twiddle_table}
      
\end{figure*}

Subsequently, we divide the ROM into $n_{PE}$ ROMs to distribute them in a way that avoids conflicts when multiple PEs attempt to access the same ROM. Noting that each pair of consecutive twiddle factors in the ROM consists of $\omega(k)$ and $\omega\left(k\pm \frac{\pi}{2}\right)$, We can further compress each ROM by a factor of 2. These twiddle factors can be derived from each other by multiplying $\omega(k)$ by $i$ or $-i$. We only store the twiddle factors of even indices in the ROM. The twiddle factors of odd indices can be easily calculated in hardware by swapping the real and imaginary parts (using MUXs) and negating the sign bit of one of them. Since the coefficient RAM and the ROMs are read simultaneously in the same clock cycle, the introduced MUXes and negation do not result in any additional delay on the critical path. Hence, this calculation does not impact the overall performance. As a result, the overall twiddle factor ROM size is reduced by a factor of four compared to the original FALCON implementation, with almost no overhead. Figure \ref{twiddle_table} illustrates an example of the proposed preprocessing and compression of the twiddle factor ROM when $n_{PE}=2$.


\subsection{Conflict-Free Scheduling}\label{Scheduling}

Memory conflicts may arise in the architecture when two PEs simultaneously request access to the same memory bank. We address this type of conflict by allocating separate memory banks for each PE to avoid contention. However, a second source of conflict is when both PE inputs are in the same memory bank. This happens because, as illustrated in Figure \ref{FFT_flow}, the distance between the indices of the two inputs of each butterfly becomes smaller for the latter stages of the $FFT^\phi$, and the earlier stages of $IFFT^\phi$. To resolve this conflict, our architecture incorporates a conflict-free scheduling algorithm (Algorithm \ref{alg:sched}) that intelligently coordinates the assignment of butterfly operations to the PEs and manages the associated memory accesses, ensuring smooth data flow and avoiding conflicts during the execution of $FFT^\phi$ or $IFFT^\phi$ operations. The algorithm we propose draws inspiration from various successful counterparts, such as the one presented in our prior work \cite{saleh2013high} and other recently introduced algorithms \cite{mu2022scalable}. It has been carefully designed and tailored specifically for the case of $FFT^\phi$/$IFFT^\phi$.

\begin{algorithm}[th!]
\caption{Conflict-Free Scheduling Algorithm}\label{alg:sched}
\begin{algorithmic}[1]
\Require $n$, $n_{PE}$, $\omega$ in ROMs, $\boldsymbol{a}$ in MEMs
\Ensure $FFT^\phi_n (\boldsymbol{a})$ in MEMs
\State $Logn \gets \log_2(n)-1$ \Comment {\textcolor{gray}{\# stages}}
\State $BT_{PE} \gets \frac{n}{4 n_{PE}}$ \Comment {\textcolor{gray}{\# operations per PE}}
\State $S_M \gets \frac{n}{4n_{PE}}$ \Comment {\textcolor{gray}{Memory size}}
\State $S_{sg} \gets \log_2(n_{PE})$ \Comment {\textcolor{gray}{Largest safe stage}}

\For{$sg = 0$ to $(Logn - 1)$}

\For{$bt_{PE} = 0$ to $(BT_{PE} - 1)$} 

    \For{$PE_i = 0$ to $(n_{PE} - 1)$}

        \State $bt \gets BT_{PE} \cdot PE_i + bt_{PE}$

        \State $Addr_0, Addr_1 \gets \text{\textbf{ MEM\_addr}}(bt_{PE}, sg)$ 

        \State $i0, i1 \gets \text{\textbf{MEM\_Select}}(sg, bt, S_M)$ 

        \State $sg_r \gets Logn - sg - 1$\Comment {\textcolor{gray}{$sg$ reverse}}
        \State $gp \gets \left\lfloor \frac{bt}{2^{sg_r}} \right\rfloor$\Comment {\textcolor{gray}{Operation group in $sg$}}

        \If{$gp$ is even \textbf{and} $sg \leq S_{sg}$} 
            \State $u \gets \text{MEM}_{i0}[Addr_0]$
            \State $v \gets \text{MEM}_{i1}[Addr_1]$
        \Else \Comment {\textcolor{gray}{Input exchange}}
            \State $v \gets \text{MEM}_{i0}[Addr_0]$
            \State $u \gets \text{MEM}_{i1}[Addr_1]$
        \EndIf

        \State $Addr2 \gets bt_{PE}$
        \State $\omega \gets \text{ROM}_{PE_i}[Addr2]$ 

        \State $[y, x] \gets \text{PE}_{PE_i}(u, v, \omega)$

        \State $EX\_Bit \gets n - sg - 2$
        \State $EX\_Flag \gets bt[EX\_Bit]$

        \If{$sg \geq S_{sg}$ \textbf{and} $EX\_Flag = 1$} 
            \State $\text{MEM}_{i0}[Addr_0] \gets y$  \Comment {\textcolor{gray}{Output exchange}}
            \State $\text{MEM}_{i1}[Addr_1] \gets x$
        \Else
            \State $\text{MEM}_{i0}[Addr_0] \gets x$
            \State $\text{MEM}_{i1}[Addr_1] \gets y$
        \EndIf

    \EndFor 
\EndFor 
\EndFor
\end{algorithmic}
\end{algorithm}

The main idea of our conflict-free scheduling algorithm is to predict whether a conflict will appear in the next stage and take action to avoid it. Let the initial inputs of $FFT^\phi_n$, denoted as $a^{(0)}_{k}=a_{k}+i\ a_{k+n/2}$, where $a_{k}\in \mathbb{Q}$ and $a^{(0)}_{k} \in \mathbb{C}$ be arranged in the memories MEM$_0$ to MEM$_{M-1}$\footnote{Actually, MEM$_i$ for $i=0,...,{M-1}$ consists of two separate memories MEM$^{real}_i$ to store the real part and MEM$^{imag}_i$ to store the imaginary part.} in the natural order, such that $k=0,...,n/2-1$. In each stage, $n/4$ butterfly operations are performed, where $bt\in{0, ..., n/4-1}$ represents the butterfly operation index in the current stage. In a specific stage $sg$, a PE$_j,\  j=0,...,n_{PE}-1$ performs $BT_{PE}$ butterfly operations from $bt=j\times n/(4\times n_{PE})$ to $bt=(j+1)\times n/(4\times n_{PE})-1$. Let $\delta^{sg}=|q-d|$ be the absolute difference between the indices $d$ and $q$ of the coefficients $a^{(sg)}_{d}$ and $a^{(sg)}_{q}$ whose butterfly operation is allocated to a specific PE in the $sg$-th stage. It can be expressed as:
\begin{equation} \label{FFT_mul2}
\delta^{sg}= \frac{n}{2^{sg+2}}.
\end{equation}

For $FFT^\phi_n$, this distance starts at $n/4$ for the first stage and is divided by two for each subsequent stage. The conflict occurs only when $a^{(sg)}_{d}$ and $a^{(sg)}_{q}$ fall in a single memory space, i.e., $\delta^{sg}$ is less than the size of the memory bank $n/(2\times M)$. We refer to the pair ($a^{(sg)}_{d}$, $a^{(sg)}_{q}$) as a Conflict-Prone (CP) pair. Hence, this conflict occurs from stage $\log_2(n_{PE})+1$ to the last stage, which we call CP stages. We define $S_{sg} = \log_2(n_{PE})$ as the largest conflict-free (safe) stage.

The procedure followed in this algorithm involves identifying the CP stages and exchanging the inputs and/or outputs of some PE operations to avoid conflicts. If the present data pair creates a hazard in the next stage, the PE outputs are exchanged. If PE inputs have been reversed previously in memory, the PE input is exchanged. This is implemented by dividing the butterfly operations into groups, with the group number varying across different stages. The current group index, $gp$, and the test for whether this stage is safe or not are used to determine if the inputs of the PEs should be exchanged. On the other hand, a specific bit, $EX\_{Bit}$, of the current butterfly operation index $bt$ is observed as an indication of whether the output of the PE needs to be exchanged in the CP stages. 

The \textbf{MEM\_Select}$(sg, bt, S_M)$ function in Algorithm \ref{alg:sched} aims to select which two memory banks are allocated to the PE, and they are different when the stages are CP or not. Because data is rearranged in memory, the algorithm needs to keep track of where the data is. In our approach, the data is moved in memory in a specific pattern to easily track the movement of data from one stage to another. To simplify the hardware implementation of this approach, we observe the pattern by which the coefficients should be exchanged in memory. As a result, the coefficient memory addresses to be accessed by the PEs are calculated in a hardware-friendly way, as illustrated in Algorithm \ref{alg:address-calculation}. For the safe stages, the PEs access the two associated memory banks with addresses $Addr_0$ and $Addr_1$, where $Addr_0$ is the index $bt_{PE}$ of the PE butterfly operation. However, for CP stages, the second address $Addr_1$ is calculated by reversing some bits of the first address $Addr_0$. The \textbf{REV}$((Addr_0, k, l))$ function reverses the bits of $Addr_0$ that start with bit $k$ and end with bit $l$. On the other hand, calculating the ROM address is very simple since a fixed single ROM is associated with each PE. In addition, the twiddle factors in ROMs are preprocessed to be accessed in the same butterfly operation order. 

\begin{algorithm}
\caption{Coefficient Memory Address Calculation}\label{alg:address-calculation}
\begin{algorithmic}[1]
\Require $bt_{PE}$, $sg$, $S_{sg}$
\Ensure PE memory addresses $Addr_0, Addr_1$
\State $Addr_0 \gets bt_{PE}$
\If{$sg \leq S_{sg}$}
    \State $Addr_1 \gets Addr_0$
\Else
    \State $k \gets S_{sg} + 1$
    \State $l \gets sg$
    \State $Addr_1 \gets \text{\textbf{Rev}}(Addr_0, k, l)$ 
\EndIf
\end{algorithmic}
\end{algorithm}

\begin{figure*}[ht]
    \centering
    \includegraphics[width=1\textwidth]{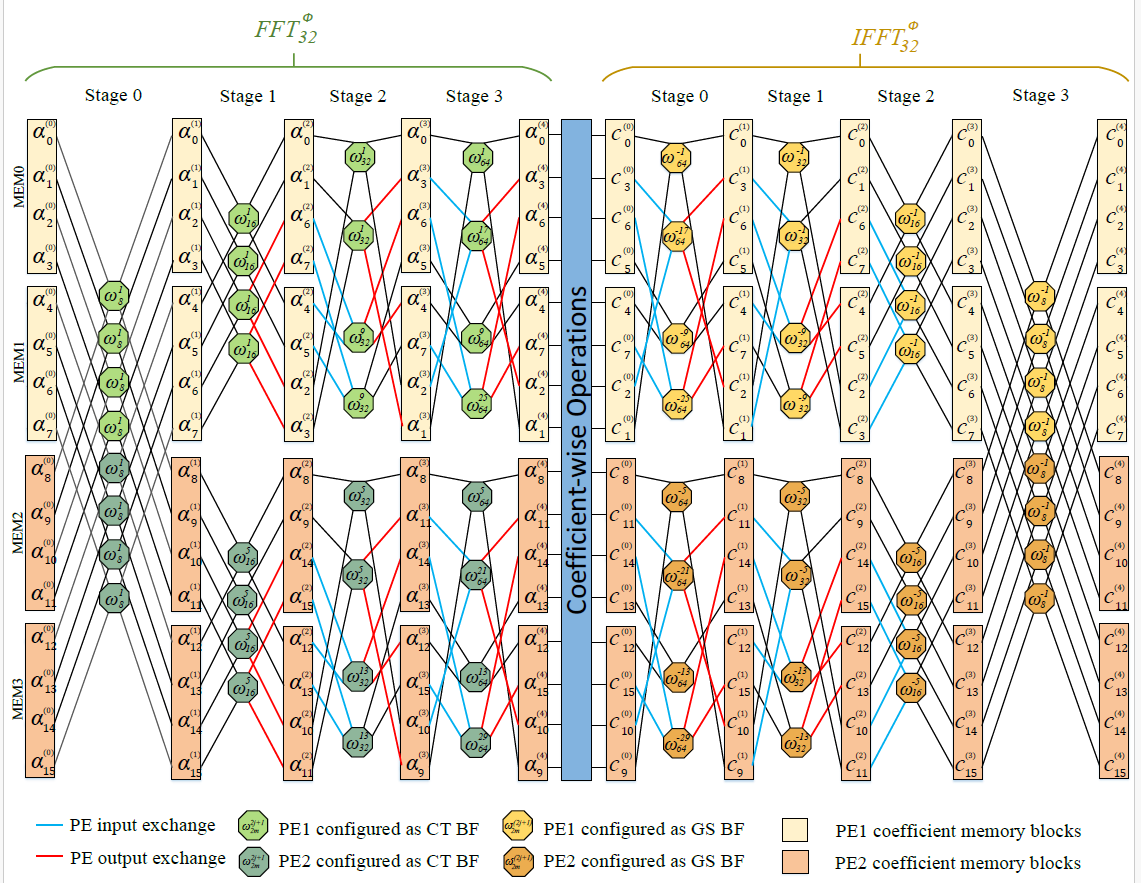}
    \caption{Conflict-free Scheduling of PE Operations for $FFT^\phi_{32}$ and $IFFT^\phi_{32}$ with $n_{PE}=2$. The figure illustrates the initial, intermediate, and final memory contents, along with the scheduling of PEs and the triggered exchange operations. In the IFFT case, the scheduling and control mirror those of the FFT.}

    \label{FFT_flow_CF}
      
\end{figure*}

Figure \ref{FFT_flow_CF} illustrates the conflict-free 32-point FFT/IFFT process using two PEs as an example. It showcases the initial, intermediate, and final memory contents, along with the scheduling of PEs and the triggered exchange operations. Notably, for the IFFT case, the scheduling and control mirror those of the FFT. This simplifies the integration of IFFT with FFT by primarily reversing the stage counter. It can be observed from Figure \ref{FFT_flow_CF} that after performing the FFT, the coefficients become disordered in memory. However, since all calculations in FALCON are coefficient-wise operations on the transformed polynomial, this disordering has no impact on the accuracy of these computations. Likewise, during the IFFT, which follows the same scheduling and control as FFT, the coefficients are reordered again to their original order. By using this conflict-free scheduling, a very simple and hardware-friendly design of the control unit is obtained.

\section{Processor Performance Evaluation}
To evaluate the proposed $FFT^\phi$/$IFFT^\phi$ processor we implement it based on the parameters listed in Table \ref{param}. Since the proposed processor is intended to serve power-constrained devices, we select $n_{PE}$ to be 2 such that two butterflies are performed in parallel. Each butterfly consists of four double-precision
FP multipliers and six double-precision FP adders/subtractors. The multipliers and adders are designed to completely follow the IEEE-754 double-precision FP data format.

\begin{table}[t]
	\caption{FFT implementation parameters}
	\centering
	\label{param}
	\resizebox{0.8\columnwidth}{!}{%
		\begin{tabular}{|cc|}
			\hline
			\multicolumn{2}{|c|}{\textbf{Processor parameters}}                         \\ \hline
			\multicolumn{1}{|c|}{$n_{PE}$}                     & 2                        \\ \hline
			\multicolumn{1}{|c|}{Largest FFT size $S_{max}$} & 1024                     \\ \hline
			\multicolumn{1}{|c|}{FALCON Security level} & I and V                      \\ \hline
			\multicolumn{2}{|c|}{\textbf{VLSI Implementation parameters}}                 \\ \hline
			\multicolumn{1}{|c|}{Process}                      & GF FD-SOI ®22 nm \\ \hline
			\multicolumn{1}{|c|}{Supply voltage}               & 0.72 V                   \\ \hline
			\multicolumn{1}{|c|}{Temperature}                  & -40C$\sim$125C           \\ \hline
			\multicolumn{1}{|c|}{CLK frequency}                & 167 MHz                  \\ \hline
			\multicolumn{1}{|c|}{SRAM number$\times$size}                     & 8 $\times$ 1KB           \\ \hline
			\multicolumn{1}{|c|}{ROM number$\times$size}                     & 2 $\times$ 2KB           \\ \hline
		\end{tabular}%
	}
\end{table}

This processor supports all security levels of FALCON since it accepts the FFT/IFFT input to be between $n= 2$ and $n=S_{max}=1024$. When $n=2$, the output of the processor is already calculated, i.e., $a^{(0)}_{0}=a_{0}+i\ a_{1}$. When $n=4$, only one of the two butterflies is used while the others are idle. Otherwise, always the two butterflies participate in the calculation simultaneously. 

\subsection{ASIC Implementation}
In the logic design stage of our processor chip, the processor is implemented using Verilog. The desired functionality of the processor is simulated using Modelsim and comprehensively tested with test vectors generated by the FALCON reference software. To keep our area and power as low as possible, the twiddle factor ROMs are implemented using look-up tables (LUTs). Our design uses hard IP blocks for the coefficient memory banks. Eight single-port $128 \times 64$bit SRAM banks with one read/write port were compiled by Synopsys Memory Compilers. These compilers generate the IP blocks and all the physical information needed for the physical design stage. We synthesized the $FFT^\phi$/$IFFT^\phi$ processor using Synopsys Design Compiler with GF FD-SOI 22 nm standard cell library.

\begin{figure}[ht]
    \centering
    \includegraphics[width=0.5\textwidth]{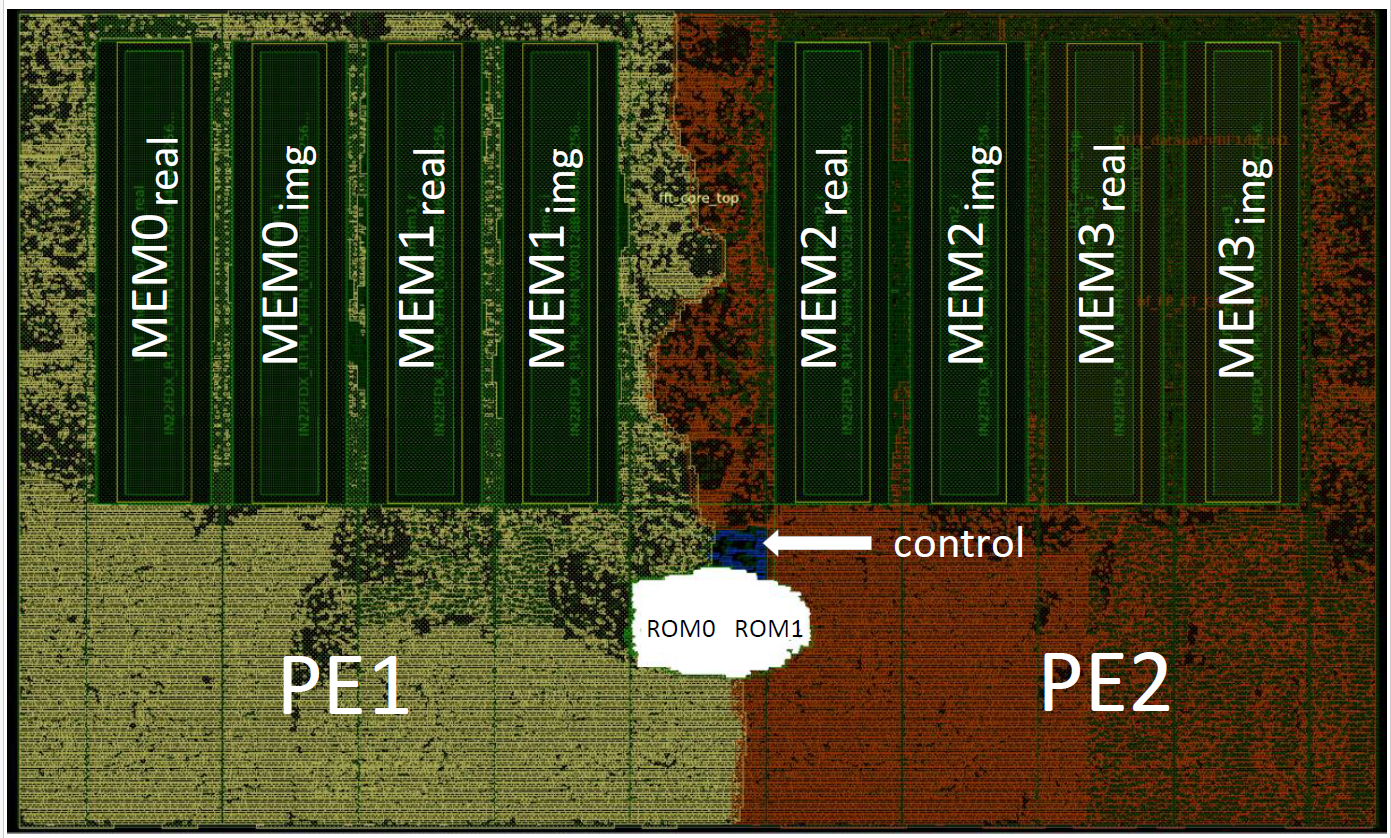}
    \caption{Floorplanning of $FFT^\phi$/$IFFT^\phi$ processor. The chip area is 0.3 mm $\times$ 0.5 mm, the clock frequency is 167 MHz, and the power consumption is 12.6 mW.}
    \label{floorplan}
      
\end{figure}

In the physical design stage, Synopsys IC Compiler II is used to perform floorplan, clock tree synthesis optimization, placement and routing optimization, and chip finishing. We targeted 6ns for the clock period. The final results show that 1) the area is 0.3 mm $\times$ 0.5 mm,  2) the clock frequency achieves 167 MHz under a wide temperature range from $- 40\degree C \text{ to } 125\degree C $, and the critical path is located in the butterfly unit in PE0, and 3) under the corner ($ 40\degree C \text{ and } 0.72 V$) the power consumption is calculated to be 12.6 mW. Figure \ref{floorplan} shows the layout of the FFT processor with sub-block annotations.

\subsubsection{Area and Power Evaluation}
\begin{figure}[ht]
    \centering
    \includegraphics[width= \linewidth]{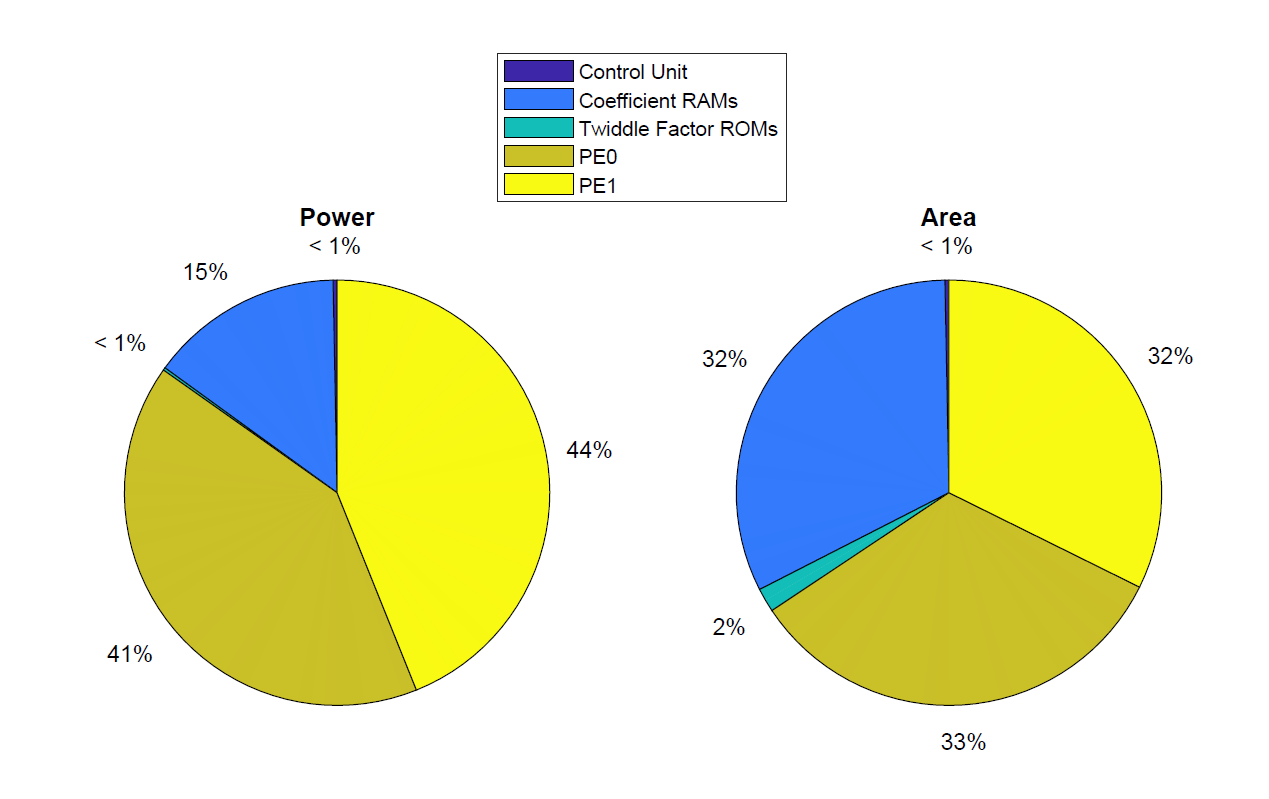}
    \caption{Area and power breakdown of the $FFT^\phi$/$IFFT^\phi$ processor (Total Area: $0.15 mm^2$, Total Power: $12.6 mW$) }
    \label{pie}
      
\end{figure}

Figure \ref{pie} shows the area and power breakdown of the $FFT^\phi$/$IFFT^\phi$ processor. The area and power needed for the control unit and ROMs are very small compared to the other components, because of the optimizations we applied; mainly the twiddle factor compression and the conflict-free scheduling. In addition, the two PEs consume the majority of the power and area of the chip, 85\% and 65\%, respectively. This is attributed to the fact that these PEs are designed to support the double-precision FP. 

\subsection{Comparisons with Related Work}
In this section, we evaluate our processor by comparing its performance to related work in the literature. First, we compare the performance of the proposed processor with a software implementation of the reference $FFT^\phi$ included in the original FALCON implementation. This $FFT^\phi$ has been implemented on a Raspberry Pi 4 with a Cortex-A72 processor and the results of this implementation are reported in \cite{nguyen2022fast}. Cortex-A72 is a high-performance and low-power processor that utilizes the ARMv8-A architecture. We compare with the reference implementation on this processor because it is an example of a relatively power-efficient processor that is suitable for embedded IoT devices \cite{becker2021neon}. At the same time, this processor has an FP unit that supports double-precision FP for a fair comparison. Table \ref{refr} illustrates the comparisons of the number of cycles and execution times for $FFT^\phi_n$ for $n$ is a power of 2 between 8 and 1024 executed on Cortex-A72 and the proposed processor. Although the clock frequency of the proposed processor is lower than that of Cortex-A72 with 1.8GHz, the calculation of $FFT^\phi_n$ for all $n$ is faster. In fact, the speed-up of our processor is from 1.1$\times$ to 2.3$\times$, and it is 1.4$\times$ on average,
for $FFT^\phi_n$ compared to the software implementation on Cortex-A72.

\begin{table*}[t]\fontsize{8pt}{12}\selectfont
	\caption{Comparison of the Cycle counts and the execution time for the implementation of $FFT^\phi_n$ with different numbers of coefficients ($n$) on Cortex-A72 and the proposed processor}
	\centering
	\label{refr}
\begin{tabular}{|c|cc|cc|}
\hline
                           & \multicolumn{2}{c|}{\textbf{Proposed}}                                          & \multicolumn{2}{c|}{\textbf{Refernce FALCON Implementation \cite{nguyen2022fast}}}           \\ \hline
\textbf{Platform}          & \multicolumn{2}{c|}{22 nm CMOS}                                        & \multicolumn{2}{c|}{ARM® Cortex®-A72}                                  \\ \hline
\textbf{Clock frequency}   & \multicolumn{2}{c|}{167MHz}                                            & \multicolumn{2}{c|}{{\color[HTML]{000000} 1.8GHz}}                     \\ \hline
\textbf{FFT size ($n$) }          & \multicolumn{1}{c|}{\textbf{Cycles}} & \textbf{Excution Time {[}ns{]}} & \multicolumn{1}{c|}{\textbf{Cycles}} & \textbf{Excution Time {[}ns{]}} \\ \hline
\textbf{8}                 & \multicolumn{1}{c|}{4}               & 24                              & \multicolumn{1}{c|}{100}             & 55.6                            \\ \hline
\textbf{16}                & \multicolumn{1}{c|}{12}              & 72                              & \multicolumn{1}{c|}{232}             & 128.9                           \\ \hline
\textbf{32}                & \multicolumn{1}{c|}{32}              & 192                             & \multicolumn{1}{c|}{516}             & 286.7                           \\ \hline
\textbf{64}                & \multicolumn{1}{c|}{80}              & 480                             & \multicolumn{1}{c|}{1,132}           & 628.9                           \\ \hline
\textbf{128}               & \multicolumn{1}{c|}{192}             & 1,152                            & \multicolumn{1}{c|}{2,529}           & 1,405                            \\ \hline
\textbf{256}               & \multicolumn{1}{c|}{448}             & 2,688                            & \multicolumn{1}{c|}{5,474}           & 3,041.1                          \\ \hline
\textbf{512}               & \multicolumn{1}{c|}{1,024}           & 6144                            & \multicolumn{1}{c|}{11,807}          & 6,559.4                          \\ \hline
\textbf{1,024}              & \multicolumn{1}{c|}{2,304}           & 13,824                           & \multicolumn{1}{c|}{27,366}          & 15,203.3                         \\ \hline
\end{tabular}

\end{table*}

Since this is the only work that custom implements $FFT^\phi$/$IFFT^\phi$ over a ring, we compare the performance of our processor with several classic FP-based FFT accelerators available in the literature. Table \ref{works} presents the performance comparison of the proposed $FFT^\phi$/$IFFT^\phi$ processor with several state-of-the-art accelerators. In this table, different accelerators use different technologies, supply voltages, FFT sizes, and data widths. Hence, the direct comparison between these processors is not fair. For fair comparison, We use normalized metrics, normalized area, and normalized power, similar to the ones proposed in \cite{beulet2014low, chen2018variable} that consider these variations. The formulae that calculate normalized area, normalized power, and normalized energy metrics with respect to our implementation are presented in (\ref{nor_area}), (\ref{nor_power}), and (\ref{nor_energy}), respectively.  
\begin{equation} \label{nor_area}
\hat{A}= \frac{A * 1000}{n*(\frac{L_m}{22})^2*(\frac{W_L}{64})},
\end{equation}
\begin{equation} \label{nor_power}
\hat{P}= \frac{P * 1000}{n*(\frac{V}{0.72})^2*(\frac{W_L}{64})},
\end{equation}
\begin{equation} \label{nor_energy}
\hat{E}= \hat{P} * t_E,
\end{equation}
where $A$ and $P $ are non-normalized values of the area and power, $t_E$ is the execution time, $n$ is the FFT size and we select it to be 1024, $L_m$ is the minimum channel length of the technology in nm normalized to ours (22 nm), $V$ is the supply voltage for each processor normalized to ours (0.72), and $W_L$ is the word size which is 64bit (double-precision FP) in our case. The normalized area, power, and energy are listed in the last three rows of Table \ref{works}. It is noticeable that our processor has the lowest normalized area and power values and, hence, it has the most power and area efficiency compared to the compared counterparts. In terms of normalized energy, our processor shows a slight increase, approximately 1.08$\times$ more than the design presented in \cite{chen2018variable}. However, this is accompanied by a substantial reduction of 7.7$\times$ in normalized area and 9.3$\times$ in normalized power. This is attributed to the higher execution time of our processor compared to \cite{chen2018variable}.        

The larger execution time of our processor reported in Table \ref{works} compared to most of the other works is attributed to the fact that our processor has more capabilities represented by its ability to perform both FFT and IFFT over the ring. In addition, several optimizations have been adopted to reduce the power and the area such as; (i) using only 2 butterfly units (PEs) to reduce the area and power overhead, (ii) using single-port instead of dual-port SRAMs which reduces the power and the area of the coefficient memory almost by the half, (however, it requires 2 cycles to read from and write to the memories which duplicate the execution time), and (iii) using double-precision FP arithmetic which is a requirement for FALCON implementation. As far as we know, the proposed processor is the only processor that performs $FFT^\phi$/$IFFT^\phi$. In addition, it has the lowest reported area and power consumption in the literature, which makes it the best fit to enable FALCON implementation on resource-constrained devices. 

\begin{table*}[t]\fontsize{8pt}{12}\selectfont
	\caption{Comparing the performance of the proposed $FFT^\phi$/$IFFT^\phi$ processor with a state-of-the-art classic FP-based FFT accelerators}
	\centering
	\label{works}
\begin{tabular}{|c|c|c|c|c|c|}
\hline
                           & Proposed             & \cite{wang2016pipelined}   & \cite{chen2018variable}   & \cite{wang2019area}                & \cite{beulet2014low}           \\ \hline
Technology/Supply voltage  & 22 nm/ 0.72V          & 65nm/1.0V  & 45nm/0.9V  & 65nm/1.0V               & 45nm/1.08 V       \\ \hline
FFT size                   & 4-1024               & 16-1024    & 4-1024     & 16-1024                 & 32–2048            \\ \hline
\# Butterfly units         & 2 radix2             & 5 radix4   & 4 radix2   & 3 radix4, 2 half radix4 & 16 radix2          \\ \hline
Data type                  & 64-bit FP            & 32-bit FP  & 32-bit FP  & 32-bit FP               & 32-bit FP          \\ \hline
Area {[}mm$^2${]}             & \textbf{0.15}                 & 1.003      & 2.4        & 0.736                   & 0.973              \\ \hline
Clock frequency            & 167 MHz              & 500MHz     & 1 GHz      & 400MHz                  & 100MHz             \\ \hline
Power [mW]             & \textbf{12.6}                 & 43.5–372.3 & 91.3       & 35.9–129.5              & 68                 \\ \hline
Execution time (1024 ) [$\micro s$] & 13.8                 & 2.03       & \textbf{1.38}       & 2.56                  & 196                \\ \hline
Run-time configurability            & \textbf{FFT/IFFT, input size} & Input size & Input size & Input size              & Input size \\ \hline

Normalized area {[}mm$^2${]}           & \textbf{0.146}                & 0.224      & 1.12       & 0.164                   & 0.227              \\ \hline
Normalized power [mW]          & \textbf{12.3}                 & 377      & 114.1      & 131.1                   & 29.5            \\ \hline
Normalized Energy [\micro J]          & 170	&	765	 & \textbf{157}	& 335	& 6,090            \\ \hline
\end{tabular}

\end{table*}
\section{Conclusion}
Considering that FFT/IFFT is the most time and power-consuming operation for FALCON PQC implementation, an optimized $FFT^\phi$/$IFFT^\phi$ processor customized for efficient FALCON implementation is presented in this paper. The proposed processor is reconfigurable at run-time to adapt to different FFT sizes (4 to 1024 points) and can perform either $FFT^\phi$ or $IFFT^\phi$ operations. Several optimization techniques are proposed in this paper, such as twiddle factor compression and conflict-free scheduling to save the power and area of the proposed processor. The ASIC implementation of the proposed $FFT^\phi$/$IFFT^\phi$ processor shows excellent performance in terms of power, area, and execution time. It achieved a speed-up compared to a software implementation of FALCON on a Cortex-A72 processor and outperformed other FP FFT accelerators in terms of power and area efficiency. This design, being the most area and energy-efficient for $FFT^\phi$/$IFFT^\phi$, is highly suitable for enabling FALCON implementation on resource-constrained devices. This processor can be used as a co-processor that accelerates the massive $FFT^\phi$/$IFFT^\phi$ operations of full implementation of FALCON PQC.

\section*{Acknowledgments}
This work was supported by the Khalifa University of Science and Technology under SOCL grant 2018-018-20 and CIRA-2020-053.

\bibliographystyle{IEEEtran}
\bibliography{IEEEabrv,references}

\begin{IEEEbiography}[{\includegraphics[width=1in,height=1.25in,clip,keepaspectratio]{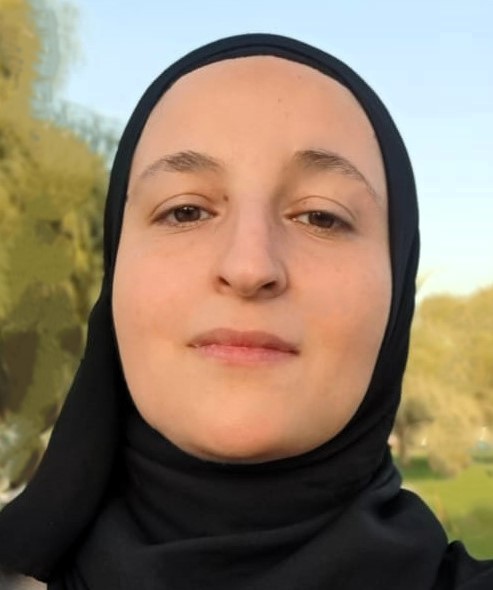}}]{Ghada Alsuhli}
earned her B.S. and M.S. in Electronics and Communication Engineering from Damascus University, Syria (2009, 2015), and completed her Ph.D. in Electronics and Communication Engineering at Cairo University, Egypt (2019). Her academic journey was enriched by roles at esteemed research centers including the National Research Center, The American University in Cairo, Egypt, and the Khalifa University SoC Center, UAE. Currently, she serves as a Post-Doctoral Researcher at Khalifa University, leading several projects focused on efficient hardware implementation for AI and post-quantum cryptography. Her research encompasses embedded systems, energy-efficient IoT solutions, edge computing, efficient hardware implementation, and AI applications for wireless communications and biomedical engineering. She is the primary author of numerous papers in reputable journals and conferences, she has also authored a book on efficient DNN hardware implementation. Her contributions extend to reviewing for esteemed journals and committee memberships for international conferences.

\end{IEEEbiography}

\begin{IEEEbiography}[{\includegraphics[width=1in,height=1.25in,clip,keepaspectratio]{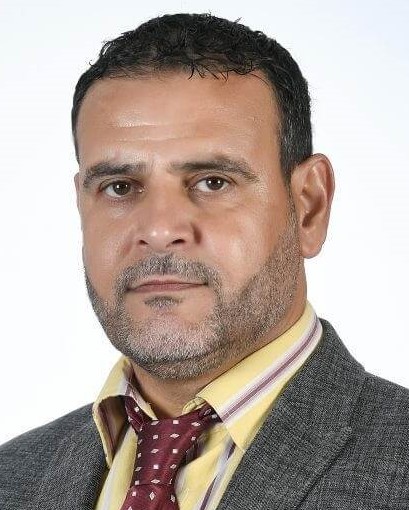}}]{Hani Saleh}
(Senior Member, IEEE) is an associate professor of  ECE at Khalifa University since 2017, he joined Khalifa on 2012. He is a co-founder of Khalifa University Research Center 2012-2018, and the System on Chip Research Center (SoC 2019-present). Hani has a total of 19 years of industrial experience in ASIC chip design, Microprocessor/Microcontroller design, DSP core design, Graphics core design and embedded systems design. Hani a Ph.D. degree in Computer Engineering from the University of Texas at Austin. Prior to joining Khalifa University he worked for many leading semiconductor design companies including Apple, Intel, AMD, Qualcomm, Synopsys, Fujitsu  and Motorola Australia. Hani has published more than 48 journal papers, more than 119 conference papers, more than 6 books and 7 book chapters. Hani research areas includes but not limited to: AI Accelerator design, Digital ASIC Design, Digital Design, Computer Architecture and Computer Arithmetic.

\end{IEEEbiography}
\begin{IEEEbiography}[{\includegraphics[width=1in,height=1.25in,clip,keepaspectratio]{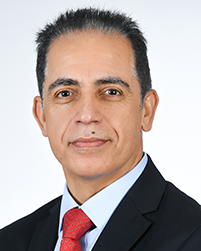}}]{Mahmoud Al-Qutayri}
(Senior Member, IEEE) is Professor of Electrical and Computer Engineering at Khalifa University (KU), UAE. He is also affiliated with KU System-on-Chip Center. Prior to joining Khalifa University, he worked at De Montfort University, UK, and the University of Bath, UK. He also worked at Philips Semiconductors, Southampton, UK. Dr. Al-Qutayri has authored or coauthored numerous technical papers in peer-reviewed journals and international conferences. His current research interests include embedded systems, in-memory computing and emerging memory technologies, energy-efficient IoT systems, efficient edge computing and artificial intelligence hardware implementation, application of AI to wireless communication systems, wireless sensor networks, and cognitive wireless networks. He received a number of awards during his education and professional career. His professional services include serving on the editorial board of some journals as well as membership of the steering, organizing, and technical program committees of a number of international conferences.

\end{IEEEbiography}

\begin{IEEEbiography}[{\includegraphics[width=1in,height=1.25in,clip,keepaspectratio]{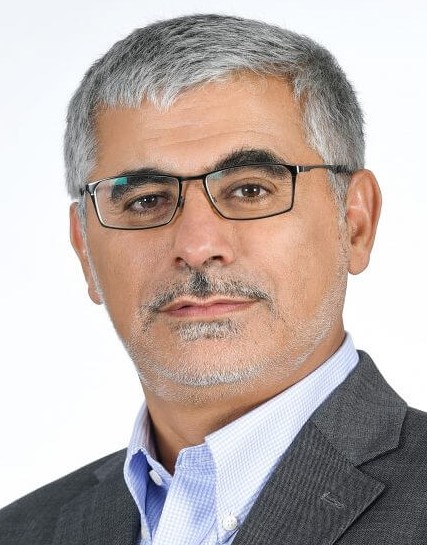}}]{Baker Mohammad}(Senior Member, IEEE)
holds a PhD in Electrical and Computer Engineering (ECE) from the University of Texas at Austin and an M.S. in ECE from Arizona State University, Tempe a B.S. degree in ECE from the University of New Mexico, Albuquerque. He is a distinguished lecturer of IEEE CAS.  Dr Mohammad is currently a professor of Electrical Engineering and Computer Science (EECS) at Khalifa University and is the director of the SOCL.  Before joining Khalifa University, Dr. Baker worked for 16 years in the US industry (Qualcomm \& Intel), designing low-power and high-performance processors. Baker’s research interests include VLSI, power-efficient computing, high-yield embedded memory, and emerging technologies such as Memristor, STTRAM, In-Memory-Computing, Hardware accelerators and power management. Dr. Baker has authored or co-authored over 200 referred journals and conference proceedings, more than three books, and over 20 U.S. patents. 
\end{IEEEbiography}

\begin{IEEEbiography}[{\includegraphics[width=1in,height=1.25in,clip,keepaspectratio]{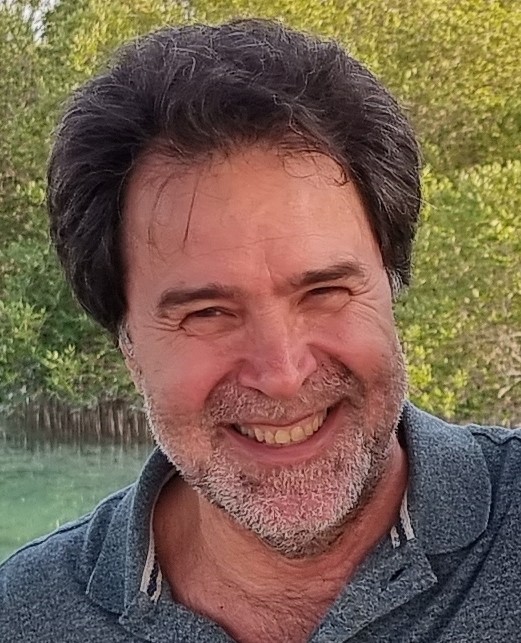}}]{Thanos Stouraitis}
(IEEE Life Fellow) is Professor of EECS at Khalifa University, UAE, and Professor Emeritus of the University Patras. He has also served on the faculties of Ohio State University, University of Florida, New York University, and University of British Columbia. He holds a Ph.D. from the University of Florida. His current research interests include AI hardware systems, signal and image processing systems, computer arithmetic, and design and architecture of optimal digital systems with emphasis on cryptographic systems. He has authored about 200 technical papers, several books and book chapters, and holds patents on DSP processor design. He received the IEEE Circuits and Systems Society Guillemin-Cauer Award. He has served as Editor or Guest Editor for numerous technical journals, as well as General Chair and/or Technical Program Committee Chair for many international conferences. He was President (2012-13) of IEEE Circuits and Systems Society.

\end{IEEEbiography}


\end{document}